\documentclass{article}
\usepackage[utf8]{inputenc}
\usepackage{amsmath}
\usepackage[a4paper, total={6in, 8in}]{geometry}
\usepackage{algorithm}
\usepackage{algpseudocode}
\usepackage{graphicx}
\usepackage{authblk}

\usepackage[sorting=none]{biblatex}
\title{Estimating the household secondary attack rate with the Incomplete Chain Binomial model}

\author[1]{Jonas C. Lindstrøm}
\author[1]{Terese Bekkevold}
\author[2]{Cathinka Halle Julin}
\author[1]{Anna Hayman Robertson}
\author[1]{Lisbeth Meyer Næss}

\affil[1]{Division of Infection Control, Norwegian Institute of Public Health, Oslo, Norway.}
\affil[2]{Division of Health Services, Norwegian Institute of Public Health, Oslo, Norway.}

\date{}

\begin{document}

\maketitle

\begin{abstract}
The Secondary Attack Rate (SAR) is a measure of how infectious a communicable disease is, and is often estimated based on studies of disease transmission in households. The Chain Binomial model is a simple model for disease outbreaks, and the final size distribution derived from it can be used to estimate the SAR using simple summary statistics. The final size distribution of the Chain Binomial model assume that the outbreaks have concluded, which in some instances may require long follow-up time. We develop a way to compute the probability distribution of the number of infected before the outbreak has concluded, which we call the Incomplete Chain Binomial distribution. We study a few theoretical properties of the model. We develop Maximum Likelihood estimation routines for inference on the SAR and explore the model by analyzing two real world data sets.   
\end{abstract}

\section{Introduction}

Chain Binomial models are a family of models for infectious diseases, and have particularly been studied in the context of household studies. In household studies a group of households are followed after a household member becomes infected (the index case), and the disease spread is then observed within the household. Understanding the dynamics of disease spread within households are important since the household is an important setting where infection can happen. Since all the individuals within the household are exposed to the infectious agent, households provide an environment where key characteristics of an infectious diseases can be studied. Such characteristics include generation time, serial interval and transmission rates, as well as factors that can influence these, such as age, crowding, or virus or bacterial sub-type.

One key parameter that can be estimated in a household study is the \emph{secondary attack rate} (SAR), sometimes also referred to as the \emph{secondary infection rate}. The household SAR is defined as the probability that the disease will spread from one infectious individual to a susceptible individual within the household. 

In practice the SAR is estimated by the proportion of household members that become infected (see for example the WHO household study protocol \cite{whoprotocol}), and the ordinary Binomial model is commonly used as the underlying statistical model. This approach is problematic, as the estimated proportion of cases in a household is not the same as the probability of an infected person infecting a susceptible household member. The two quantities are in fact two distinct concepts that should not be conflated, and the proportion of infected household members is more accurately referred to as the \emph{final attack rate} (FAR) \cite{10.1371/journal.pcbi.1008601}. The final attack rate will for example include situations where individuals that escaped infection by the index case instead became infected by the secondary cases or tertiary cases. The final number of cases will therefore be expected to be greater than the SAR. The final attack rate will also depend on the size of the household, in addition to the SAR. A larger household will have a greater expected FAR than a smaller household, even if the SAR is the same in both households \cite{10.1371/journal.pcbi.1008601}. If this is not properly addressed, it will seriously affect the estimates and conclusions of a study. This will also have consequences for comparing studies conducted in different countries, as the household sizes can vary a lot. For example is the average household size in Germany 2.1 (in 2011), while in Pakistan it is 6.8 (in 2013) \cite{unreport}

While the ordinary Binomial model with cluster adjusted standard errors could be a suitable model for the FAR \cite{10.1371/journal.pcbi.1008601}, other approaches should be used to make inferences on the SAR. The Chain Binomial model is one such candidate model. In the Chain Binomial model the disease spread is modelled as occurring in discrete time steps referred to as \emph{generations}. The number of new cases in each generation is modelled using ordinary Binomial models, and by \emph{chaining} together Binomial models for each generation we get a more realistic model that accounts for the spread of disease from secondary cases to tertiary cases, and so on. The SAR is the only parameter of the model. 

Methods for efficiently computing the probability distribution of the final outbreak size of the Chain Binomial model have been known for some time \cite{LUDWIG1975}. This makes the use of the Chain Binomial an attractive alternative to the ordinary Binomial model for inference, since the only data needed are the initial number of susceptibles and the final number of cases, which are the same as needed for the ordinary Binomial model. A further improvement of the Chain Binomial over the ordinary Binomial is that it can easily account for the situation where there is more than one index case. Outbreaks where there are more than one index case are often excluded from the analysis in household studies.

A crucial assumption that must be made when using the final outbreak distribution of the Chain Binomial is that the observation time of a household is long enough to actually observe the entire outbreak. In a meta-analysis of household studies on Covid-19, 16 of 54 studies had follow-up times that were short ($< 14$ days) or did not report it \cite{Madewell}. Short follow-up times will be a bigger problem in large households. In the most extreme case, the spread within the household might be such that in each generation only 1 person become infected, thus the observation time of a large household, with for example 9 initial suceptibles, will be 9 generations. If the observation time is only 3 generations, for example, the "final" Chain Binomial model might not be suitable. 

The aim of this paper is to develop a way to use the Chain Binomial model for estimating the SAR for both when the outbreak is completely observed and when it is incompletely observed in time. We estimate the SAR both in the case of \emph{i.i.d.} households and when the attack rate might depend on some household-level varying factors, using a GLM-like modelling approach. We have implemented a R package called \texttt{chainbinomial} that makes these methods available. We will also study some properties of the Chain Binomial model, such as the bias that might occur when assuming a final distribution size when the outbreak is still ongoing within households. In order to explore how the model performs on real world data, we have used data from our recent SARS-CoV-2 household study \cite{microorganisms9112371} as well as historical data from a common cold survey in the UK \cite{Brimblecombe_1958} \cite{Heasman_reid_1961}.


\section{The Chain Binomial Model}

Consider a household with $N$ members, where one or more household members acquire an infection from outside the household at a time point 0. We denote the number of initial cases $I_0$, and the remaining non-infected and susceptible household members as $S_0$, so that $I_0 + S_0 = N$.

Let $\alpha \in (0,1)$ be the secondary attack rate (SAR), which is the  probability that the disease will spread from one infectious individual to a susceptible one within the household. We assume that there are no individual differences in susceptibility or infectiousness, and that all household members have the same rates of contact with each other (homogeneous mixing), so this parameter does not differ between pairs of individuals in the household. This is the parameter we are interested in making inferences on. 

To model the spread of infections within the household over time, we consider time in discrete generations, denoted $g$, with the initial infections happening at $g=0$. In each generation, the remaining suceptibles can become infected by any of the infected household members, but they are not considered infectious until the next generation. We do not consider the possibility of any additional infections from outside the household other than the initial $I_0$. The infectious household members are assumed to be infectious only for a single generation, after which they are considered immune and do not contribute more to disease spread in the household.

From one generation to the next, the number of new infected household members is modelled using a binomial model.  Let $I_g$ be the number of infectious in generation $g$, and $S_g$ be the number of remaining susceptible persons in the household at generation $g$. Then 

\begin{equation} \label{cb_model1}
P(I_{g+1} | I_{g}, S_g. \alpha) = \binom{S_g}{I_{g+1}} \pi_{g+1}^{I_{g+1}} (1-\pi_{g+1})^{S_g - I_{g+1}}
\end{equation}

where $\pi_{g+1}$ is the per-person probability that a remaining susceptible in generation $g$ will be infected by any of the $I_g$ infected in generation $g$. This 
is the probability of having at least one binomial "success" of $I_g$ independent trials with probability $\alpha$. Hence 

\begin{equation}
\pi_{g+1} =  1 - \binom{I_g}{0}\alpha^0(1-\alpha)^{I_g}  = 1 - (1-\alpha)^{I_{g}} \,.
\end{equation}

When $\pi_{g+1}$ is modelled in this way, the Chain Binomial model is sometimes referred to as the Reed-Frost model \cite{Becker1989}. Other approaches are possible, including a completely general approach that have a separate $\pi$ for each combination of $g$ and $I_g$ \cite{Becker1989}, or as a constant parameter that do not depend on $I_g$, which is referred to as the Greenwood model. However, these models do not have the property that the parameter is easily interpreted as the secondary attack rate in the specific sense we want to use here.

\subsection{Scenario probabilities}

Next we can consider the probability of a specific scenario, where a scenario is a given number of infected in each generation, up to generation $r$ denoted $i_0 \rightarrow i_1 \rightarrow ... \rightarrow i_r$. We can compute the probability of a specific scenario by repeated use (or \emph{chaining}) of Equation \ref{cb_model1}. 

\begin{equation} \label{cb_model_chain_prob}
p(i_0 \rightarrow i_1 \rightarrow ... \rightarrow i_r; S_0, \alpha) = \prod_{g=0}^{r-1} P(I_{g+1} = i_{g+1} | I_{g} = i_{g}, S_{g} = s_{g}, \alpha) \,,
\end{equation}

where $s_{g} = S_0 - \sum_{t=1}^{g-1} i_t$ is the number of remaining susceptibles in generation $g$. An outbreak will conclude if all $S_0$ sucecptibles become infected, or if there is a generation where no one becomes infected.

The modelling of time into discrete generations, where no one is infectious for more than one time step, and the secondary cases are only infectious in the next generation, might be unrealistic for many diseases. Diseases that best fits this assumption are diseases where the latent, pre-infectious, period is longer than the infectious period. Applying the Chain Binomial model to diseases that do not fit this description might not be a serious obstacle as long as we are not focusing on specific chains, but instead concentrate on the outbreaks sizes, as we will explore in the next section.

\subsection{The final and incompletely observed outbreak sizes}

While it is possible to analyze data on specific scenarios using Equation \ref{cb_model_chain_prob}, we will instead consider the case where we only have the size of the outbreak as counts of the number of infected household members, either after the entire outbreak has played out, or at some earlier time point.

An interesting method for computing the probability distribution of the final size of the outbreak is given in \cite{LUDWIG1975}. The method is interesting since it does not rely directly on studying the scenario probabilities such as in Equation \ref{cb_model_chain_prob}. Instead the final size distribution can be computed by considering how the probability distribution changes when a new person is added to the pool of susceptibles, who then might or might not become infected. This bypasses the need to consider the timing of the infections, but makes the approach unsuitable for computing the probability distribution after a given number of generations. 

To compute the probability for an observed total count $I_d = \sum^d I$ after a given number of generations $d$, we can instead use Equation \ref{cb_model_chain_prob} and sum over all possible scenarios with lengths from 1 to $d$ that gives $I_d$ infected individuals. 

\begin{equation} \label{incomplete_cb}
p(I_d,  S_0, \alpha) = \sum p(i_0 \rightarrow i_1 \rightarrow ... \rightarrow i_d;  S_0, \alpha) \,.
\end{equation}

We refer to this as the Incomplete Chain Binomial probability of $I_d$. Summing over all possible scenarios is basically a brute force approach to the problem, but this is not a problem for a modern computer for the typical household sizes we will consider here. Pseudocode for an algorithm for finding all scenarios to sum over is presented in Algorithm \ref{alg:allscenarios}. The number of scenarios that need to be considered for computing $p(I_d)$ is related to the number of \emph{compositions} from 1 to $d$ parts of the number $I_d$ \cite{enwiki:1153982709}, and is given by 

\begin{equation}
    nscenarios(I_d, d) = \sum_{j=1}^{d} \binom{I_d - 1}{j-1} \,.
\end{equation}

Briefly explained, the algorithm starts with the scenario $i_0 \rightarrow 1 \rightarrow 1 \rightarrow ... \rightarrow I_d - d + 1$, which is the extreme case of 1 infected in each generation, with the remaining infections happening in the $d$-th generation. We can ignore the first generation $i_0$ since it is actually observed, and only consider scenarios from generation 1 and onward. To generate the next scenario we look at the second-last generation, and increase the number of infected there by 1, so that $i_{d-1} = 2$, while also setting $i_{d} = 0$. If the scenario vector sums to $I_d$ then we have found the next scenario, otherwise we increase the number of infected in the subsequent generations by 1 until the desired sum is reached. Then the process is repeated by increasing the number of infected in earlier generations and adding the remaining infections needed in later generations until we have found all scenarios. When $I_d = 0$, $I_d = 1$, or $d = 1$, the number of generations needed to reach $I_d$ is 1, and only 1 scenario is possible, and the resulting scenario is trivially given.

Note that this approach can also be used to compute the final size distribution. To do that all we need is to sum over all scenarios of lengths from 1 to $S_0$ generations, since this is the longest an outbreak can last, which happens in the extreme case of only 1 new infected in each generation until everyone has become infected. 

\begin{algorithm}
\caption{Find all possible scenarios giving x infected in g generations}
\label{alg:allscenarios}
\begin{algorithmic}
\Require $x \geq 0$ \Comment{Target sum}
\Require $d \geq 1$ \Comment{Number of generations}
\State $nScenarios \gets nscenarios(x, g) $ \Comment{Compute the number of scenarios}
\State $allScenarios \gets matrix[d, nScenarios]$

\If{$x = 0$} 
\State $allScenarios[, 1] \gets 0$

\Return{$allScenarios$}

\EndIf

\If{$x = 1$} 
\State $allScenarios[1, 1] \gets 1$

\Return{$allScenarios$}

\EndIf

\If{$d = 1$} 
\State $allScenarios[, 1] \gets x$

\Return{$allScenarios$}

\EndIf

\State $initScenario \gets [1, 1, 1, ..., (x - d + 1)]$
\State $nextScenario \gets initScenario$
\State $depth \gets d$
\For{$i = 2$ to $nScenarios$}

\State $nextScenario[depth:d] \gets 0$ 
\State $depth  \gets depth - 1$
\State $nextScenario[depth] \gets nextScenario[depth] + 1$ 

\While{$sum(nextScenario) < x$}
\State $depth  \gets depth + 1$

\If{$depth = d$}
\State $nextScenario[depth] \gets x - sum(nextScenario[1:depth])$
\Else
\State $nextScenario[depth] \gets nextScenario[depth] + 1$
\EndIf
\EndWhile

\State $allScenarios[, i] \gets nextScenario$

\EndFor

\Return{$allScenarios$}

\end{algorithmic}
\end{algorithm}

\subsection{Estimation and inference}

We will use Maximum Likelihood to estimate the SAR $\alpha$ using the probability mass functions of the final and incomplete Chain Binomial models. If we have data from $n$ independently observed households, with known known observation time of $d_i$ generations for household $i$, the likelihood function is

\begin{equation}
    \mathcal{L}(\alpha | I, S_0, I_0, d) = \prod_{i=1}^n p_i(I_{d_i}, S_{0i}, I_{0i};\alpha) \,,
\end{equation}

where $p_i$ is the probability of the observed data for household $i$, given $\alpha$. We optimize the likelihood function directly using numerical methods. As the incomplete Chain Binomial distribution function is a sum of individual chain probabilities, we could adapt the approach in \cite{Becker1989} for computational efficiency, but for practical purposes numerical optimization is fast enough, and also works well for the final size distribution. 

We can also let the SAR $\alpha$ depend on a set of predictors $x$, in a manner similar to the Generalized Linear Models framework. Let $\alpha_i$, the SAR for household $i$, depend on a set of $K$ predictors $x_{ik}$, for $k \in 1, .., K$, in the following way

\begin{equation}
    f(\alpha_i) = \beta_0 + \sum_{k=1}^K \beta_k x_{ik} \,,
\end{equation}

with $\beta_0$ being the intercept, and $\beta_k$ be the coefficient of the linear part of the model. The function $f(\cdot)$ is the link function, which transform the linear prediction. Since valid values of $\alpha$ is in the $[0,1]$ range, the logit link is a natural link function to use. For certain simple models, for example if there is only one binary or categorical predictor, log and identity link functions can be considered in order to make parameter interpretation easier. Estimation of the $\beta$ parameters is also done using Maximum Likelihood, with numerical optimization. 

For both the \emph{i.i.d.} case and the prediction modelling case, standard errors are computed using the inverse of the second derivative (or the Hessian matrix) of the likelihood function at the maximum, as is standard procedure. Confidence intervals for the \emph{i.i.d.} case can be computed using Wilks Theorem \cite{wilks1938}, or using the normal distribution centered around the point estimate and the standard error as the standard deviation. 

\begin{figure}
    \centering
    \includegraphics[width=0.9\linewidth]{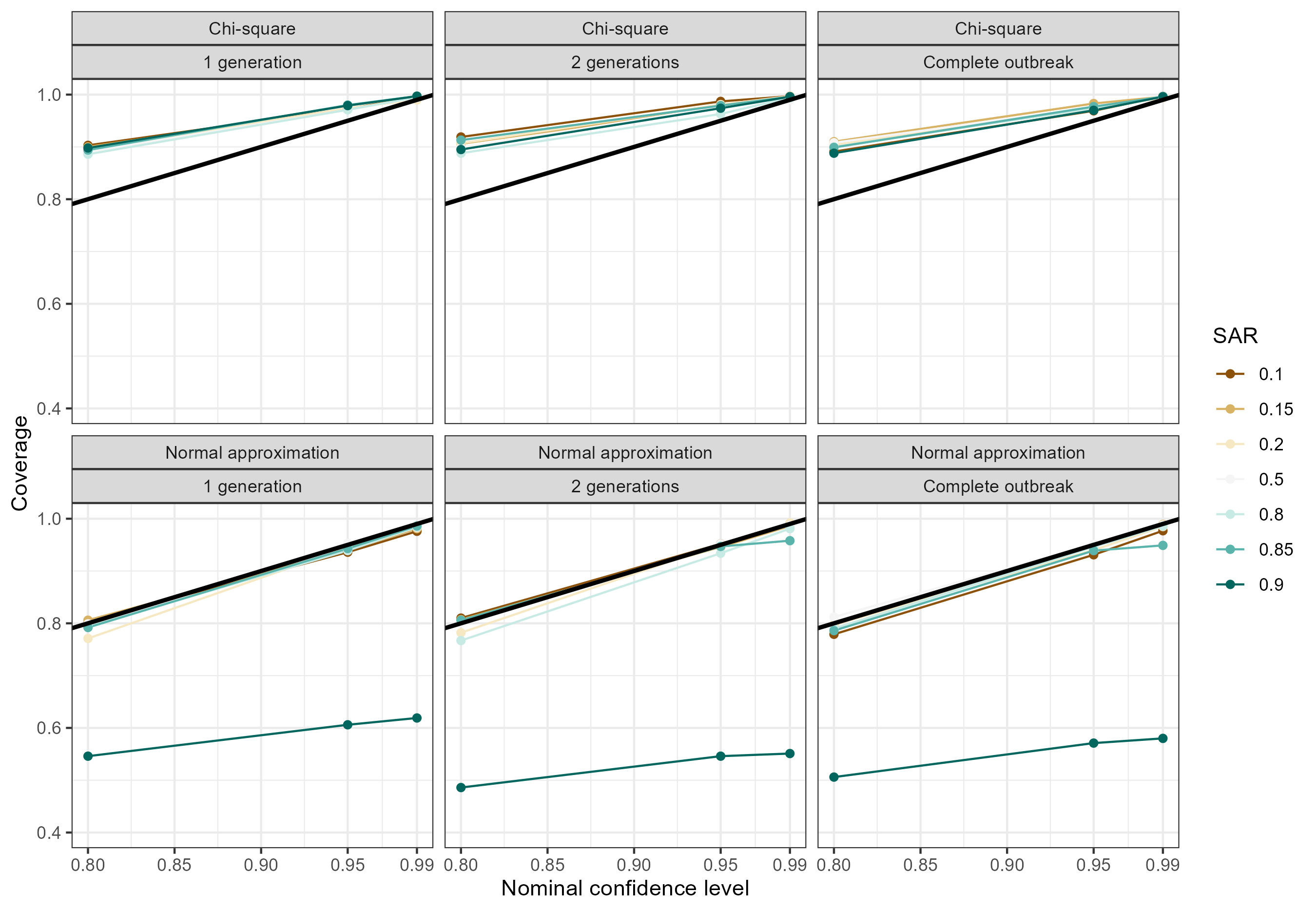}
    \caption{Results from simulation study, comparing the realized coverage of two methods for constructing confidence intervals. 100 household outbreaks were simulated with the observation time of 1 and 2 generations, and completely observed, for different SAR values (lines). The black line represent the line of equality. The Chi-square method based on Wilks theorem (top row) shows a slightly more conservative coverage than the nominal coverage, with 80\% intervals having a 90\% realized coverage. The method based on the normal distribution (bottom row) shows a slightly lower coverage than the nominal coverage, except when SAR = 0.9, where the coverage is severely misscalibrated. }
    \label{fig:coverage}
\end{figure}

We did a simulation study to compare the two approaches. We simulated outbreaks in $n = 10, 20, 50$ and $100$ households, with SAR varying between 0.1 and 0.9, for 1 and 2 generations, in addition to the complete outbreak. 1000 replications were simulated for each setup. The sizes of the households were simulated to reflect household sizes of a typical western country. Figure \ref{fig:coverage} shows the estimated coverage for $n = 100$ for 80\%, 95\% and 99\% nominal coverage.

In summary we found that the Wilks construction was preferable. Standard errors for the SAR were often impossible to compute when the SAR was $\geq 0.8$ for sample sizes (number of households) less than 50. We found that the reason for this was that for large values of the SAR there is a high probability that all household members become infected. For example is the probability that 4 out of 4 susceptibles become infected greater than 0.99 when $SAR = 0.8$. When the sample size is low it is likely that all household members in all households become infected which results in a Maximum Likelihood estimate for the SAR of 1, which rules out the possibility that a household members escapes infection. 

The confidence intervals based on the normal distribution approach was severely miss-calibrated when the SAR was 0.9, with 95\% intervals showing smaller than 60\% coverage, even for large number of households ($n=100)$. Confidence intervals based on Wilks Theorem could always be computed, but had a coverage that were slightly greater than the nominal coverage, with the 80\% interval having a 90\% coverage. 

For the analyses where SAR depends on predictors, we use the normal approximation approach.

\section{Comparison of the final vs incomplete chain binomial distributions}

\subsection{The Incomplete Chain Binomial model converges to the final size distribution}

\begin{figure}
    \centering
    \includegraphics[width=\textwidth,height=\textheight,keepaspectratio]{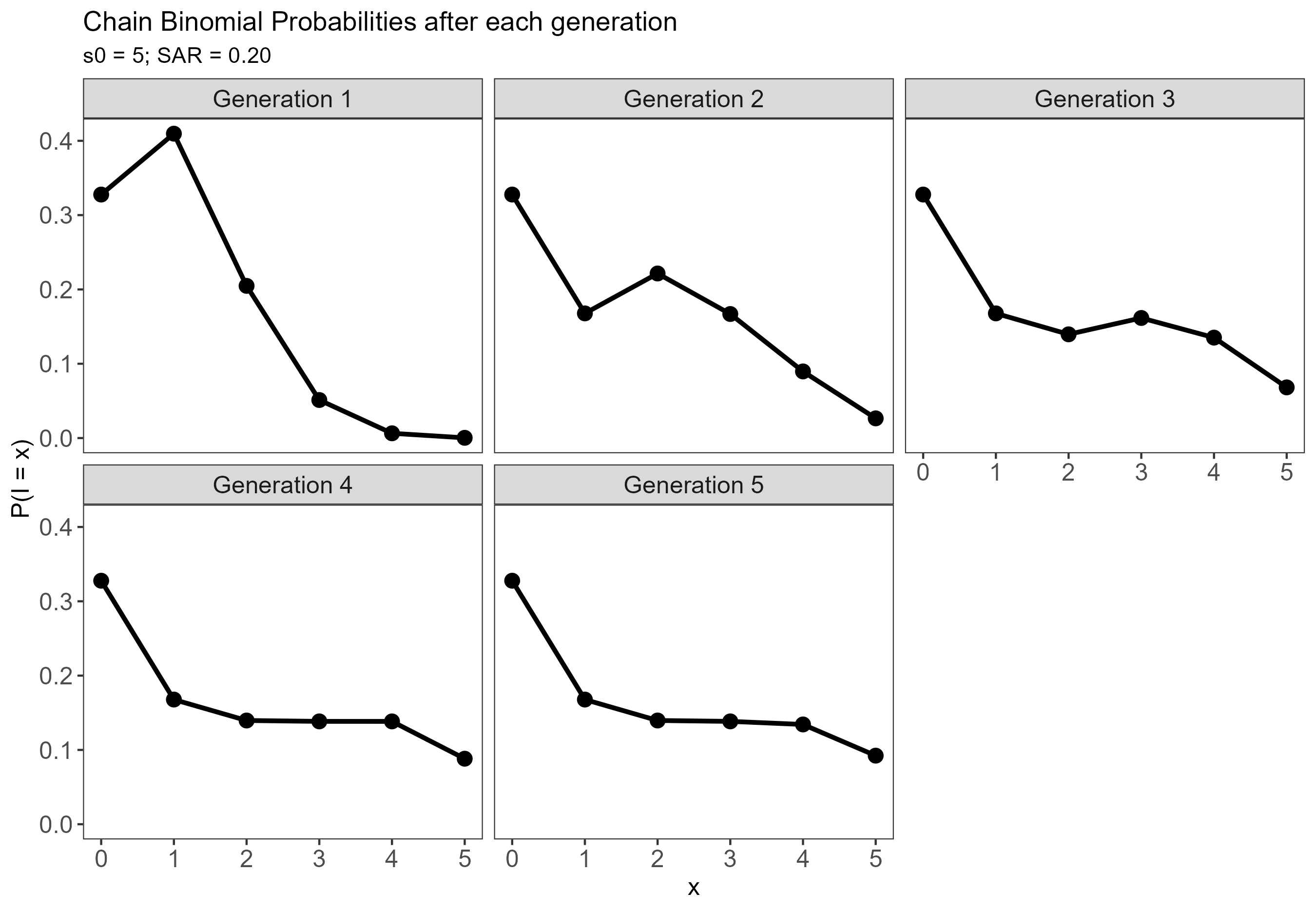}
    \caption{The Incomplete Chain Binomial distribution for $S_0 = 5$ initial susceptibles, $I_0 = 1$ initial infected, SAR $\alpha = 0.2$, after each generation. The 5th generation corresponds to the final size distribution, while the 1st generation distribution is an ordinary Binomial distribution. }
    \label{fig:pltcbsar20}
\end{figure}

Figure \ref{fig:pltcbsar20} illustrates how the Chain Binomial probability distribution changes over time for a set of parameters. As the number of generations increases, the Incomplete Chain Binomial probability distribution converges to the final size distribution. The reason for this is that the longest time it would take for exactly $I = i$ infections to occur is $i$ generations, for $i > 0$. This occurs in the case when only one person is infected in each generation, \emph{i.e.} the chain is $1 \rightarrow 1 \rightarrow ... \rightarrow 1$. For the case of $I = 0$, one generation has to be observed in order to observe that no one becomes infected. Together this means that $p(I=i, g=d+1) = p(I=i, g=d)$ for $i < d$, $d>1$.

\subsection{The SAR estimated from the final size distribution is biased when the outbreak has not concluded}

If an outbreak is observed only for a short time, and the outbreak is still ongoing when the observations time ends, using the final size distribution will be inappropriate and inference for the SAR will be biased. To investigate the extent of this bias we will compare the estimates of SAR that is obtained when using the final size distribution when the true distribution is an Incomplete Chain Binomial after $g$ generations, for a given $S_0$ and $I_0$ and SAR.

We can compare the final and incomplete distributions using the The Kullback–Leibler divergence. Let $p_d(x, \alpha)$ be the incomplete chain binomial probability of $x = \tilde{I_d}$ infected after $d$ generations, with a true SAR $\alpha$. Then let $q(x, \hat{\alpha})$ be the final size probability of $x = I$ infected with SAR $\hat{\alpha}$. To avoid clutter we drop the $S_0$ and $I_0$ from the notation, since we always will consider these to be the same in the comparisons of the final and incomplete probability distributions.

The Kullback–Leibler divergence is given as

\begin{equation}
    KL(\alpha, \hat{\alpha}) = \sum_{x = 0}^{s_0} p_d(x, \alpha) log(\frac{p_d(x, \alpha)}{q(x, \hat{\alpha})})
\end{equation}

For a given true SAR $\alpha$, we can find the best approximating SAR  $\hat{\alpha}$ when using the final size distribution when the true data comes from an Incomplete Chain Binomial model by minimizing the Kullback–Leibler divergence. This $\hat{\alpha}$ is the expected estimate when using the final size distribution, and the difference between $\hat{\alpha}$ and $\alpha$ is the bias. 

\begin{figure}[H]
    \centering
    \includegraphics[width=\textwidth,height=\textheight,keepaspectratio]{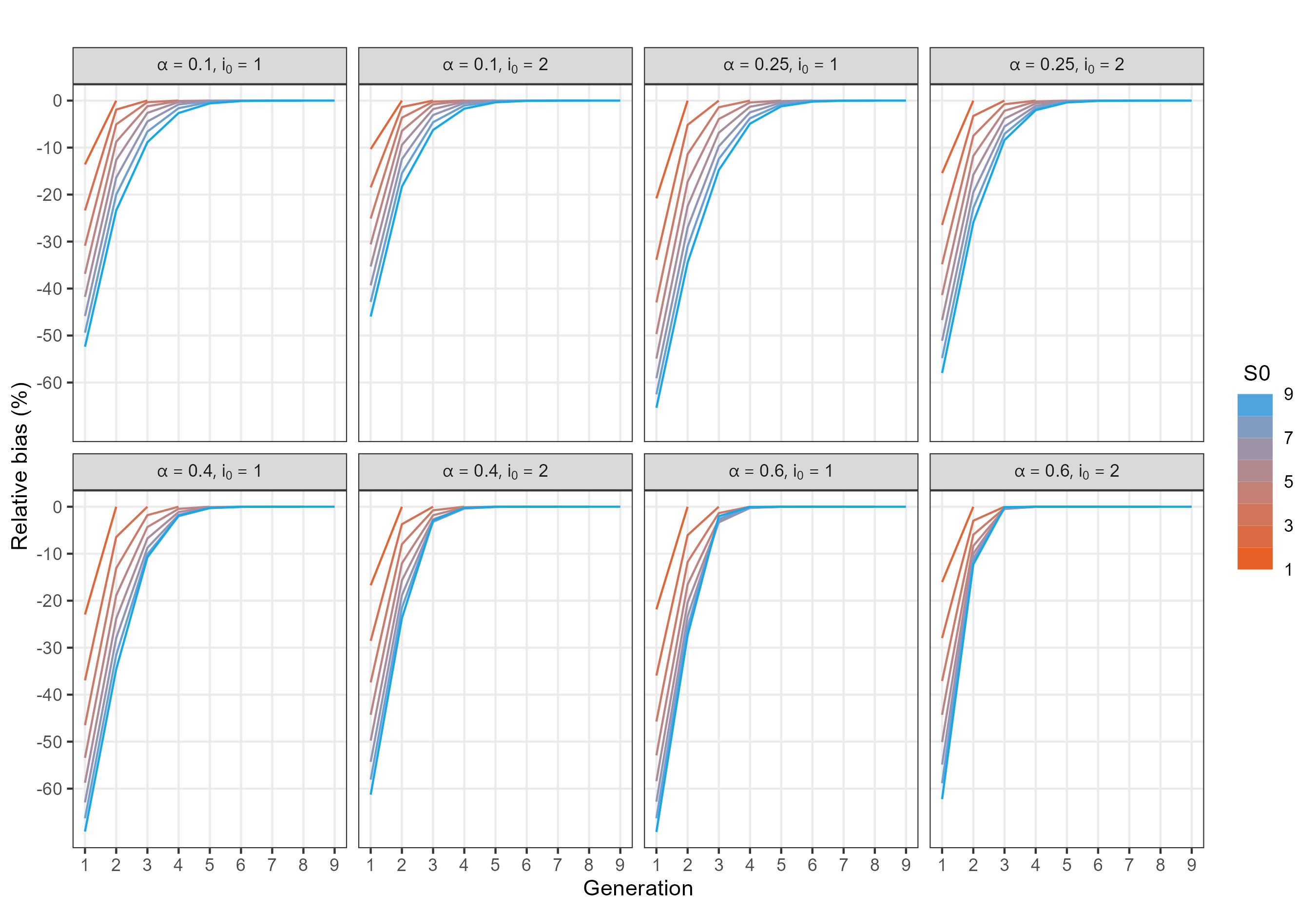}
    \caption{The relative bias in SAR when using the final size distribution when the outbreak is incompletely observed. Each line shows the bias (y-axis) as the generations unfold (x-axis) for a given number of initial susceptibles ($S_0$).}
    \label{fig:pltbias}
\end{figure}

The relative bias of $\hat{\alpha}$ is shown in figure \ref{fig:pltbias} for different values of the true SAR $\alpha$, number of initial susceptibles $S_0$ and infected $I_0$ for different observation times in generations. The final size distribution will be expected to systematically underestimate the SAR when the final outbreak has not been completely observed, and this will be more severe the fewer generations have been observed, and the greater number of initial susceptibles there are. However, for all scenarios the bias will be negligible from 5 generations and onwards.

\section{Data analysis}

\subsection{Common cold data}

To illustrate the use of the incomplete Chain Binomial model in an analysis we will use historical data from common cold survey in London \cite{Brimblecombe_1958}, which studied households consisting of two parents and three children over several seasons. The data were later studied in detail by Heasman and Reid \cite{Heasman_reid_1961}, who classified each case in an outbreak to a generation, hence giving us data that fits neatly into the Chain Binomial framework. The data set consists of chain data scenarios on 756 outbreaks, and 92 of these have two index cases. 

\begin{figure}[H]
    \centering
    \includegraphics[width=0.80\linewidth]{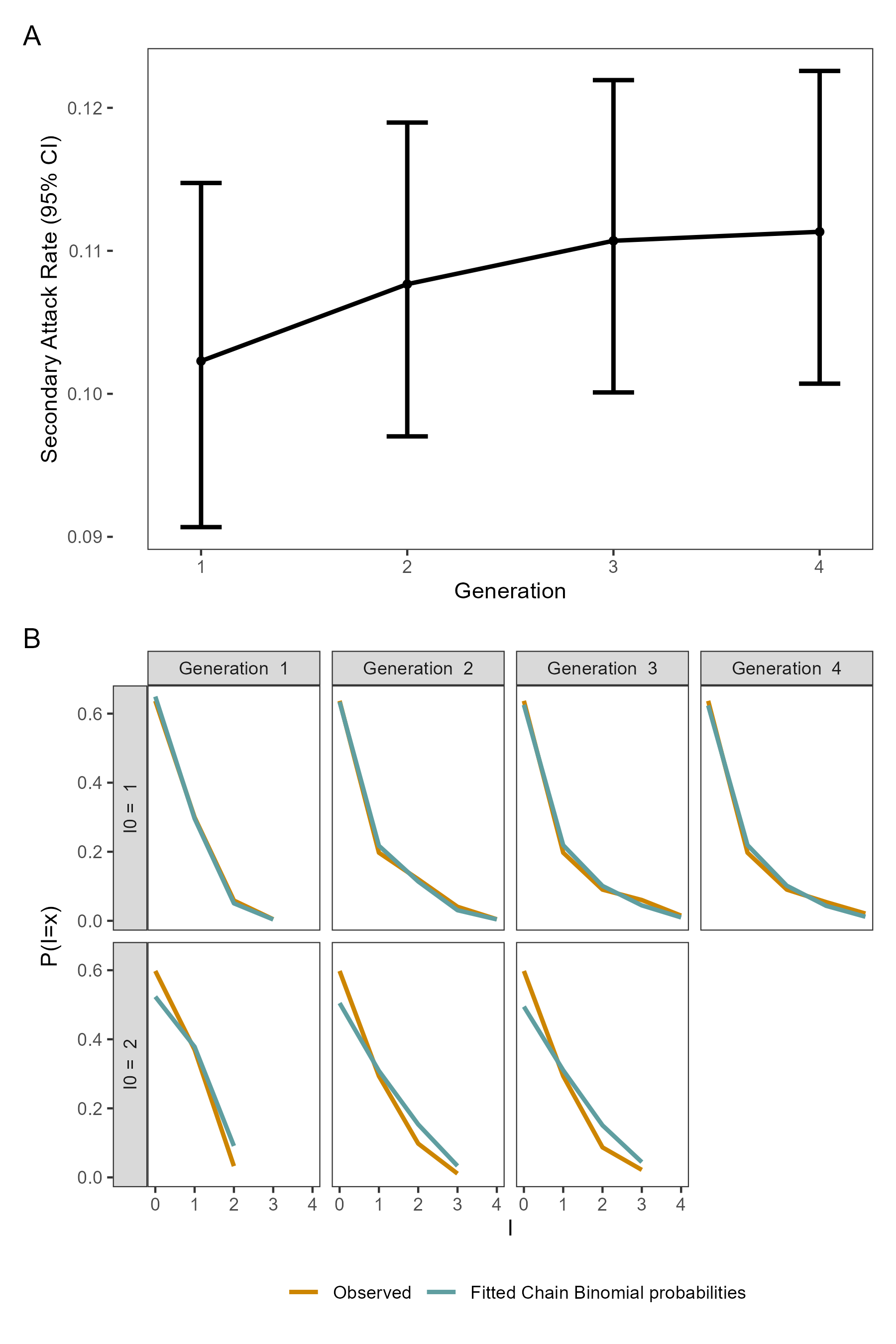}
    \caption{Analysis of Common Cold data of household of size 5. A: Estimates of SAR after each generation. B: Observed frequencies of number of cases for one index case (top row) and two index cases (bottom row) and fitted chain binomial probabilities, after each generation. }
    \label{fig:common_cold}
\end{figure}

Using the final outbreak sizes we estimate the secondary attack rate to 0.11 (95 \% CI: 0.10 - 0.12). Estimates of the secondary attack rate using the incomplete chains up to a given generation are essentially the same, with the SAR estimated from only the first generation of cases being the smallest (0.102, 95 \% CI: 0.091  0.12). Estimates and confidence intervals based on data up to a given generation for all generations are shown in Figure \ref{fig:common_cold} A, and shows that the estimates based on different generations differ less than the estimated uncertainty implied by the confidence intervals.  

In Figure \ref{fig:common_cold} B the observed frequencies of the outbreak sizes are shown after each generation, together with the fitted Chain Binomial probabilities. It shows a near perfect fit for the households with a single index case. A somewhat poorer fit is seen for the outbreaks where there were two index cases. In the original analysis of the chain data \cite{Heasman_reid_1961} the households with two index cases were analyzed separately, giving a somewhat smaller estimate of the SAR (0.081). 

\subsection{The CoronaHouse study}

The CoronaHouse study was conducted from May 2020 to April 2021 in Norway, in order to study the infectiousness of the SARS-CoV-2 virus \cite{microorganisms9112371}. 70 households, with 216 participants were recruited after one household member had tested positive for the SARS-CoV-2. Household members were followed up with regular testing for up to 42 days. The study period covered two phases of the epidemic, where the original virus variant (non-Variant of Concern, non-VOC) went from being the dominant to being replaced by the more infectious B.1.1.7 alpha variant. During the study period infected persons were ordered to isolate at home while other household members were ordered to quarantine at home and make minimal contact outside the household. As a consequence it was unlikely that household members other than the index cases were infected outside the household, which was also supported by sequencing data showing that all household members were infected by the same variant. 

In our re-analysis we included only households where all household members were eligible and agreed to participate. We excluded households where sequence data on virus variant was lacking, or were infected with other variants of concern (the Beta variant). We included 3 households with more than one index case, which where excluded in the original analysis, and one person who tested positive more than 14 days after the index case. 52 households with a total of 166 household members were included in our new analysis. The largest households had 6 members, which means that the maximum number of generations we were able to model is 5. The data for the final outbreak sizes is given in Table \ref{tab:koronahusdata}. 

The estimate of the SAR for the non-VOC based on the final size Chain Binomial model is 0.28 (95\% CI: 0.19 - 0.36). The original analysis estimated the FAR to be 0.43 (95\% CI 0.29 – 0.58), which is similar to the FAR implied by the Chain Binomial model, which is  0.40 (assuming SAR = 0.28, one initial infected and 3 initial susceptibles, since median household size was 4 in the non-VOC group).

For the B.1.1.7 alpha variant we estimated the SAR to be 0.61 (95\% CI: 0.43 - 0.79).  The original analysis estimated the FAR to be 0.78 (95\% CI: 0.49 - 0.93), which also matches the FAR implied by the Chain Binomial (FAR = 0.76, assuming SAR = 0.61, one initial infected and 2 initial susceptibles, since median household size was 3 in this group).

To compare the SAR for the two variants, we fitted a prediction model with variant as predictor, with identity link. From this model the estimated difference in SAR between the two variants was 0.33 (95\% CI: 0.14 - 0.53).

To get a sense of how the incomplete Chain Binomial model would work for these data had the study been stopped at an earlier time, we created three data sets that emulated early stopping by including only the test results at day 5, day 10, and day 15 after the index case tested positive. For these three data sets, plus the complete data set at the end of the study, we fitted five models, each assuming a known number of generations, from 1 to 5 generations. Each model had virus variant as predictor. 

\begin{figure}[H]
    \centering
    \includegraphics[width=0.7\linewidth]{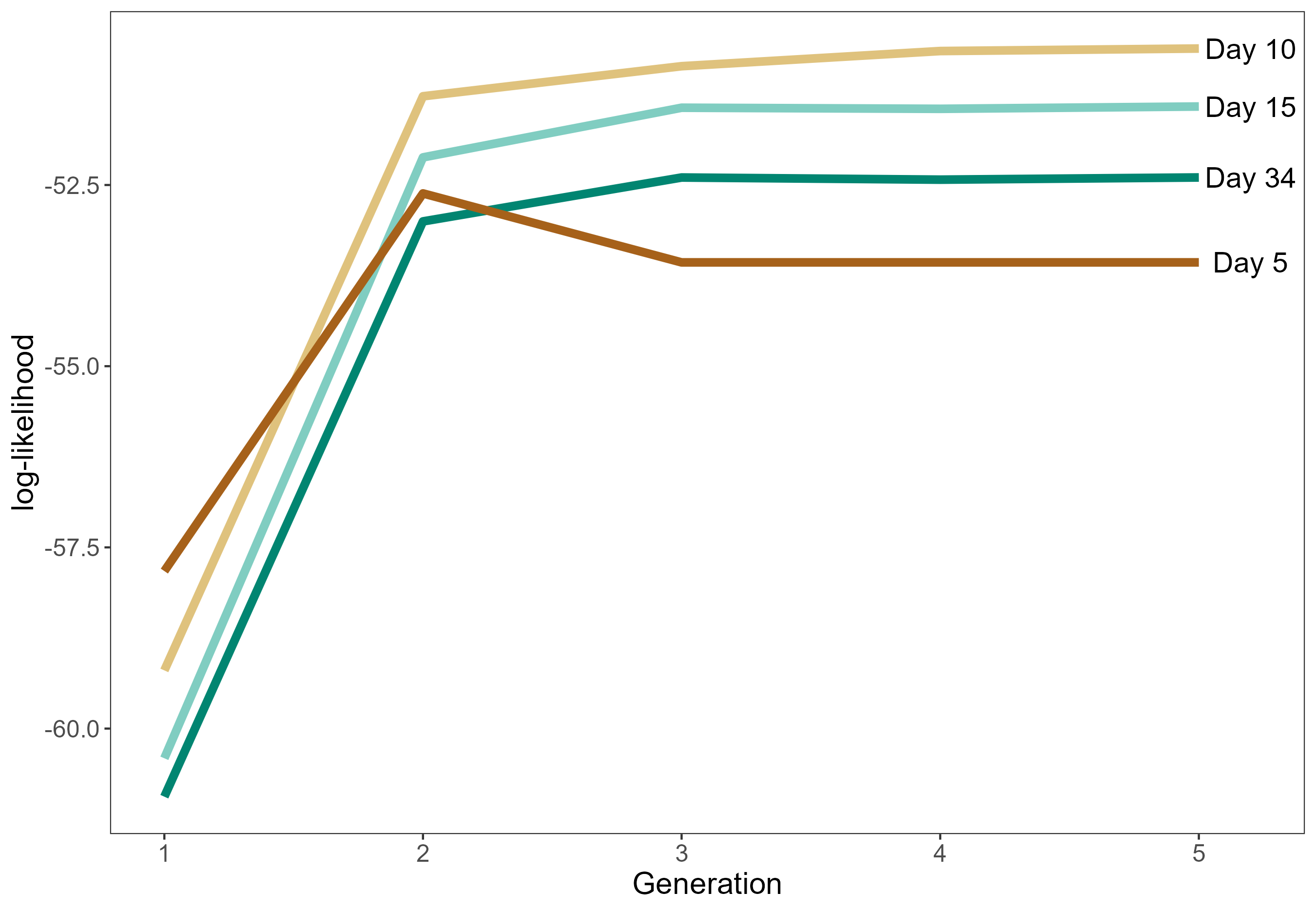}
    \caption{Comparing models fitted to CoronaHouse data truncated after 5, 10, 15 days, plus the final data (day 34). Each line corresponds to one of the truncated data sets. Model fit (log-likelihood) is shown on the y axis, while the x axis is the number of generations assumed in the model.}
    \label{fig:fitgendays}
\end{figure}

Figure \ref{fig:fitgendays} shows the log-likelihood of the fitted models for comparison. The final size distribution gave the best fit for all data sets, except for when the data were truncated after 5 days. The 5-day data set had best fit when it was assumed that two generations had been observed. For all data sets the worst fit was when only 1 generation was assumed, which corresponds to an ordinary Binomial model.

The model had in general better fit to the data that were truncated after 10 and 15 days than the final data, even when assuming the final size model, which is a bit unexpected. This could indicate that some of the infections that occurred later during the study period was from outside the household, since the were only 10 days of isolation and quarantining after the initial testing/contact. After day 10 there were however only three additional persons from three different households who tested positive. Other explanations could also apply, for example that the model does not perfectly capture the dynamics of the spread.

\begin{table}[ht]
\centering
\begin{tabular}{lc|rrrrr|rr|rr}
  \hline
Variant &  Household size & \multicolumn{5}{c|}{$I_0 = 1$}  & \multicolumn{2}{c|}{$I_0 = 2$}  & $I_0 = 3$ \\ 

 \hline
 &  & 0 & 1 & 2 & 3 & 5 & 0  & 2 & 0 \\ 
  \hline
non-VOC & 2 &   8 &   7 &   - &   - &   - &   - &   - & -  \\ 
 & 3 &   2 &   3 &   3 &   - &  - &   0 &   - & -  \\ 
 & 4 &   5 &   0 &   1 &   1 &   - &   1 &   0 & 1  \\ 
 & 5 &   2 &   1 &   1 &   0 &   - &   0 &   0 & 0  \\ 
 & 6 &   0 &   0 &   0 &   1 &   1 &   0 &   0 & 0  \\ \hline
B.1.1.7 alpha & 2 &   1 &   5 &   - &   - &   - &   - &   - & -  \\ 
 & 4 &   1 &   1 &   0 &   5 &   - &   0 &   1 & 0  \\ 
 \hline
\end{tabular}

\caption{Data from from the CoronaHouse study. The cells indicates the number of households of the given household size (rows) and number of secondary cases (columns), for both the non-VOC and alpha variants, and for different number of index cases. The $S_0$ will be the household size minus $I_0$.}
\label{tab:koronahusdata}
\end{table}

\newpage

\section{Discussion}

In this paper we have developed an approach for estimating the secondary attack rate from household studies that have an observation period that can be shorter than the entire outbreak period. This approach relies mainly on aggregated data per household, such as the number of infected, number of index cases, and number of susceptible individuals. This is a strength since the method does not need data on the timing of the infections, and neither does it depend on knowing the serial interval or generation time, which may not be available for new emerging viruses. Since we also do not need individual level data, data can easily be shared while preserving anonymity. 

We have investigated some consequences of wrongly assuming the the final size distribution when the observation time is short. Based on theoretical calculations we found that after 5 generations there is little bias in using the final size distribution, even for larger households. This knowledge could be useful for the design of future household studies. 

We analyzed two real-world data sets, one historical data set on common cold that has been studied before in the context of Chain Binomial models, and data set from a recent Norwegian household study on Covid-19 transmission. 

The historical common cold data had already been classified into chains suitable for a detailed Chain Binomial analysis, and we found that the Chain Binomial model fit the data well, both when the final outbreak sizes was considered, and when the Incomplete model was used with the truncated chains as data. 

In our analysis of Covid-19 data we had access to the timing of the test results, and could therefore mimic incomplete outbreak data. In the analysis we explored the model fit when treating the number of generations as a tuning parameter that was kept fixed while the SAR parameter was estimated. In the analyses of the data set truncated at various follow-up times we found that assuming only one generation gave the poorest fit to the data, while assuming two or more generations gave a better fit. Assuming only one generation is equivalent to assuming an ordinary Binomial model with different parameters to account for the number of index cases. 

The \emph{generations} in the Chain Binomial context does not in general correspond to other epidemiological quantities such as generation time and serial interval. The Chain Binomial model assumes that the generations are discrete and separate in time, which does not realistically applies to Covid-19. Therefore the generation should only be considered a tuning parameter, and we stress that the incomplete Chain Binomial distribution should in general not be used to estimate the generation time. 

The \texttt{chainbinomial} R package is available at  \texttt{github.com/opisthokonta/chainbinomial} and will be submitted to CRAN in the near future. 

\newpage


\printbibliography

\end{document}